\documentclass[prd,preprintnumbers,showpacs,twocolumn,nofootinbib,notitlepage,longbibliography,superscriptaddress]{revtex4-1}
\usepackage{bbold}
\usepackage{amsmath, amssymb,braket}
\usepackage{color}
\usepackage{xcolor}
\usepackage{float}
\usepackage{graphicx}
\usepackage{subfigure}
\usepackage{natbib}
\usepackage{adjustbox}
\usepackage{xspace}
\usepackage[colorlinks=true,citecolor=blue,urlcolor=blue]{hyperref}

\def\apj{\rm{ApJ}}                 
\def\apjl{\rm{ApJ}}                

\def\mnras{\rm{MNRAS}}             

\def\Msun{M_\odot} 
\def\prd{\rm{Phys.~Rev.~D}}        
\def\nat{\rm{Nature}}              

\newcommand{\avg}[1]{\left< #1 \right>}

\newcommand\orcid[1]{\href{http://orcid.org/#1}{\adjustbox{trim={-.15\width} {0\height} {-.15\width} {0\height},clip}{\includegraphics[height=10pt]{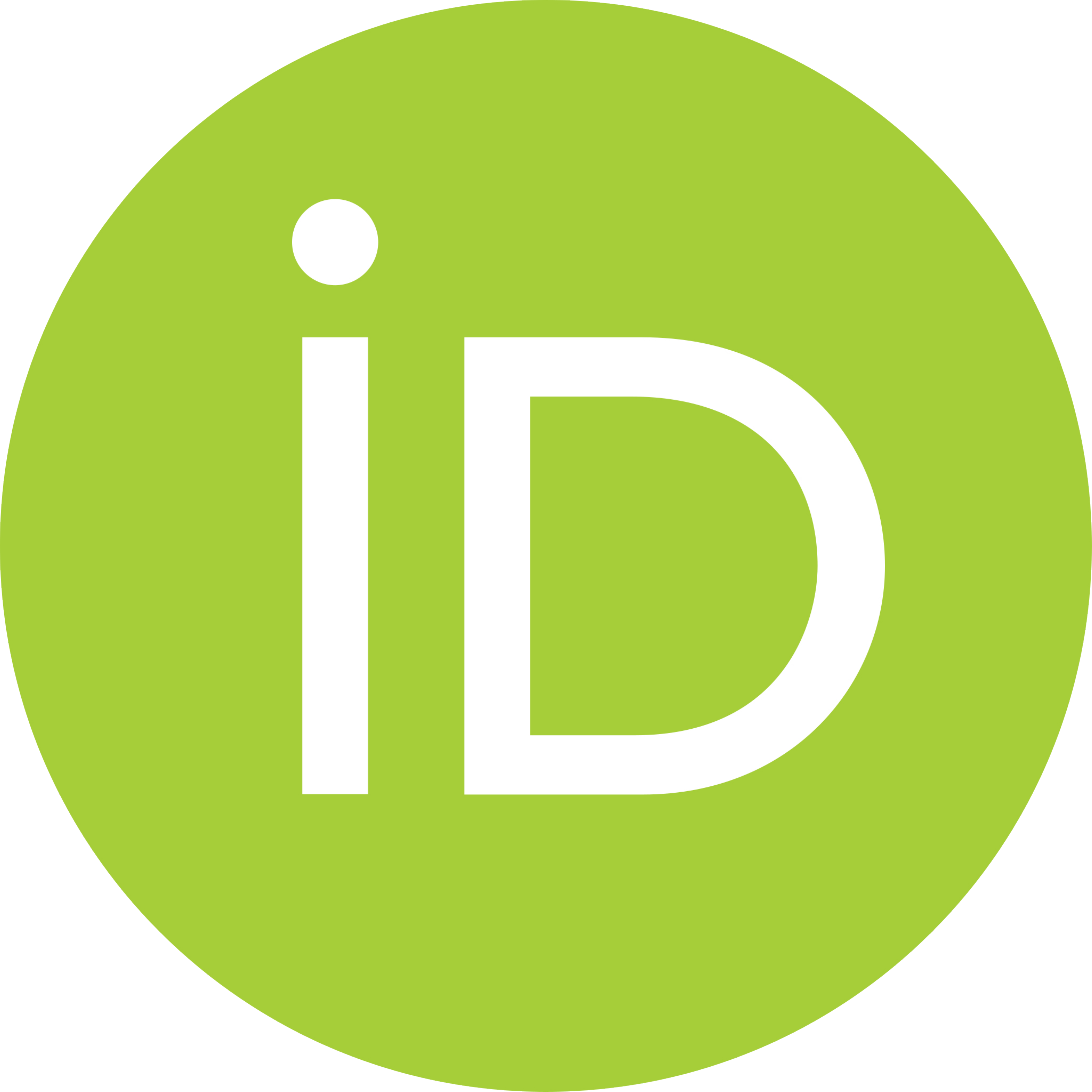}}}}
\newcommand{\n}{\nonumber \\}
\newcommand{\dbar}{d\hspace*{-0.08em}\bar{}\hspace*{0.1em}}
\newcommand{\thesan}{\textsc{thesan}\xspace}

\begin{document}

\preprint{MIT-CTP/5856}

\title{Supersizing hydrodynamical simulations of reionization using perturbative techniques}

\author{Wenzer Qin}
\affiliation{Department of Physics, Massachusetts Institute of Technology, Cambridge, MA 02139, USA}
\affiliation{Center for Theoretical Physics, Massachusetts Institute of Technology, Cambridge, Massachusetts 02139, USA}
\affiliation{Center for Cosmology and Particle Physics, Department of Physics, New York University, New York, NY 10003, USA}

\author{Katelin Schutz}
\affiliation{Department of Physics \& Trottier Space Institute, McGill University, Montr\'{e}al, QC H3A 2T8, Canada}

\author{Olivia Rosenstein}
\affiliation{Department of Physics, Massachusetts Institute of Technology, Cambridge, MA 02139, USA}
\affiliation{Department of Physics, \'Ecole Normale Sup\'erieure Paris-Saclay, Gif-Sur-Yvette, France}

\author{Stephanie O'Neil}
\affiliation{Department of Physics, Massachusetts Institute of Technology, Cambridge, MA 02139, USA}
\affiliation{Kavli Institute for Astrophysics and Space Research, Massachusetts Institute of Technology, Cambridge, MA 02139, USA}
\affiliation{Department of Physics \& Astronomy, University of Pennsylvania, Philadelphia, PA 19104, USA}
\affiliation{Department of Physics, Princeton University, Princeton, NJ 08544, USA}

\author{Mark Vogelsberger}
\affiliation{Department of Physics, Massachusetts Institute of Technology, Cambridge, MA 02139, USA}
\affiliation{Kavli Institute for Astrophysics and Space Research, Massachusetts Institute of Technology, Cambridge, MA 02139, USA}

\begin{abstract}
\noindent We show that perturbative techniques inspired by effective field theory (EFT) can be used to ``paint on'' the large-scale 21~cm field during reionization using only the underlying linear density field. It is therefore possible to enlarge or ``supersize" hydrodynamical simulations at low resolution, on scales that are larger than the nonlinear scale of the 21~cm field. In particular, the EFT provides a mapping between the linear density field and the 21~cm field. We show that this mapping can be reliably extracted from relatively small simulation volumes using the \textsc{thesan} suite of simulations, which have a comoving volume of $(95.5~\mathrm{Mpc})^3$. Specifically, we show that if we fit the EFT coefficients in a small $\sim5$\% sub-volume of the simulation, we can predict the 21~cm field to within $\mathcal{O}(10\%)$ accuracy in the rest of the simulation given only the linear density field. 
We show that our technique is robust to different models of dark matter and differences in the sub-grid reionization modeling.
\end{abstract}

\maketitle

\section{Introduction}
The study of 21~cm cosmology has emerged as a powerful tool for probing the early universe, particularly during the epoch of reionization (EoR)~\cite{Furlanetto:2006jb,2012RPPh...75h6901P}. By observing the redshifted 21~cm line emitted or absorbed by neutral hydrogen, it is possible to map the hydrogen's spatial distribution, which offers insights into the formation and evolution of cosmic structure. 
The EoR also provides a unique window onto the complex astrophysical interplay between the formation of the first stars and galaxies and the thermal history of the intergalactic medium (IGM). As observational techniques improve, 21~cm observations are poised to significantly enhance our understanding of the astrophysics and cosmology at play during the EoR~\cite{Liu:2019awk}. 
Meanwhile, there has been a concerted effort to develop theoretical techniques that can account for the intricacies of the reionization process, including analytic perturbative methods~\cite{2007MNRAS.375..324Z,2015PhRvD..91h3015M,2019MNRAS.487.3050H,McQuinn:2018zwa,21cmEFT,Obuljen:2022cjo,Pourtsidou:2022gsb,Sailer:2022vqx}, semi-analytic and effective models~\cite{2018ApJ...860...55R,2019ApJ...876...56R,Mirocha:2022,2013ApJ...776...81B,2022ApJ...927..186T,2012Natur.487...70V,2014MNRAS.437L..36F,2011MNRAS.411..955M,2020JOSS....5.2582M,Baradaran:2024jlh,Cruz:2024fsv,Munoz:2023kkg,Schaeffer:2024qcz}, and full hydrodynamic simulations~\cite{gnedin1997reionization,ciardi2003simulating,McQuinn:2006et, iliev2006simulating,trac2007radiative,gnedin2014cosmic,pawlik2017aurora}. 

\begin{figure*}
    \centering
    \includegraphics[width=\textwidth]{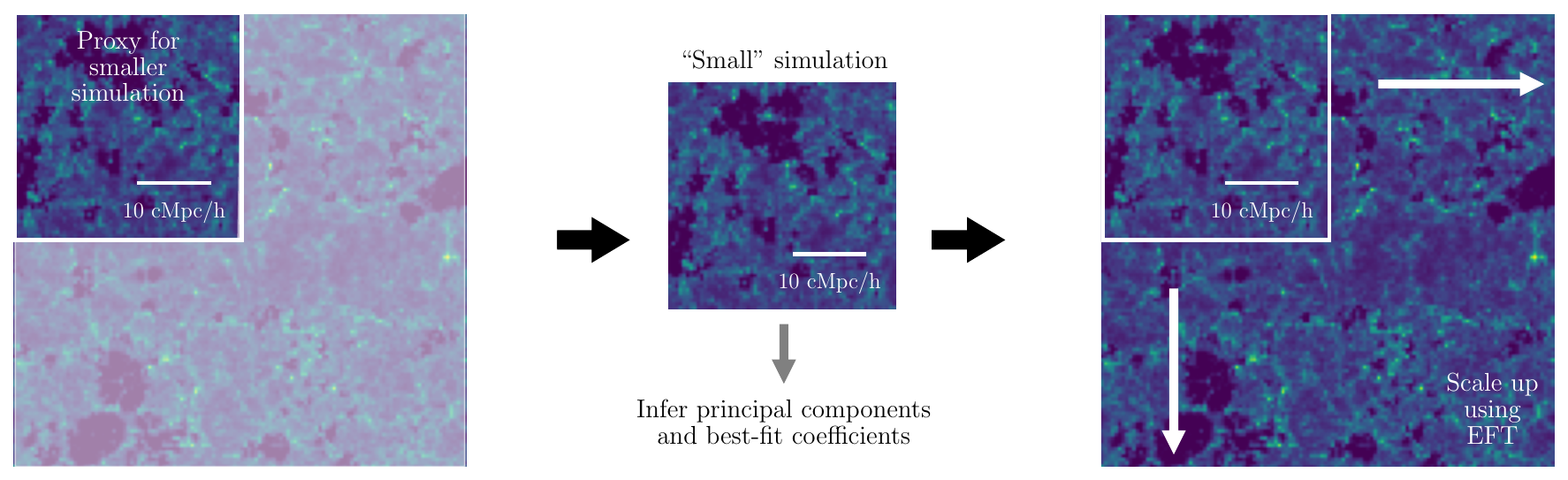}
    \vspace{-0.6cm}
    \caption{
    Diagram illustrating the use of simulation subvolumes in this work.
    We pare down the \thesan simulation to a smaller box to serve as a proxy for a simulation run at a smaller volume.
    We determine the principal components of the effective bias expansion and their best fit coefficients from the ``small" simulation and apply these to the original-size linear density field to see how closely we recover the 21\,cm signal at large scales.
    Note that the third panel shows an idealized scenario meant to illustrate how the procedure works, and does not show the actual results which are instead displayed in Fig.~\ref{fig:big_slice}.
    }
    \vspace{-0.2cm}\label{fig:diagram}
\end{figure*}

There are benefits and drawbacks to the various theoretical approaches under development. 
Notably, while state-of-the-art hydrodynamic simulations can capture the largest range of physical effects that impact the EoR, they are also extremely computationally expensive to run, particularly in large simulation volumes. 
The computational expense of large simulations is at odds with the fact that the signal is most detectable on large scales. 
At present, the strongest upper limits on the 21~cm power spectrum at $z \sim 6-10$ come from the Hydrogen Epoch of Reionization Array (HERA)~\cite{DeBoer:2016tnn} and correspond to wavenumbers of $k\sim 0.1$~Mpc$^{-1}$~\cite{2023ApJ...945..124H}, while the \textsc{thesan} simulations have a box size of 95.5~Mpc, corresponding to a similar minimum wavenumber, $k\sim 0.1$ Mpc$^{-1}$~\cite{thesan1,thesan2,thesan3}. 
There are therefore very few modes in simulations to compare against observation. 
In fact, the instantaneous field of view of HERA at 150~MHz spans around 1.4 comoving Gpc in the transverse direction. 
As such, hydrodynamical simulations are currently much too small to be useful in validating the analysis pipelines for various EoR observatories. 
The computational expense of fully state-of-the-art simulations becomes especially apparent when considering the possibility of exploring different cosmological parameters, sub-grid physics, and dark matter models.  
In order to make a positive detection of the EoR, which is orders of magnitude fainter than foregrounds, it is crucial to have an end-to-end way of testing for even tiny amounts of erroneous signal injection or loss~\cite{Cheng:2018osq, 2022MNRAS.516.5588G, Pascua:2024cmy}. 

In this work, we develop a simulation-calibrated method for painting the 21~cm field onto the linear density field on large scales using techniques inspired by effective field theory (EFT). This method can be viewed as a compromise between doing a full simulation and using a semi-analytic model: it can capture the complex, multi-scale physics of a simulation while also being parameterized by a small set of coefficients and spectral ``shapes'' in Fourier space.  
In comparison to large-volume hydrodynamical simulations, EFT-based methods require far less computational resources by virtue of providing an analytic description of cosmological fields on large scales. In the mildly nonlinear regime, one can use EFT and related methods to systematically incorporate nonlinear effects order by order in perturbation theory. For instance, it is possible to capture the effects of feedback from physics on small, non-perturbative scales to accurately describe larger scales of interest (analogous to renormalization~\cite{Assassi:2014fva}).
In particular, we employ the EFT-based description of 21\,cm fluctuations developed in Refs.~\cite{21cmEFT, McQuinn:2018zwa}, which is capable of describing redshift-space distortions, and which was shown to be able to describe simulations run with a variety of underlying physics assumptions. 

Our proposed method for generating the large-scale 21\,cm signal is depicted schematically in Fig.~\ref{fig:diagram} and can be summarized as:
\begin{enumerate}
    \item run a relatively small-volume simulation with some set of sub-grid physics parameters and an underlying cosmological model;
    \item fit EFT coefficients, which describe the transformation between the linear density field and the 21~cm field, to the simulation at the field level for modes that are larger than the nonlinear scale;
    \item generate a random realization of the linear density field at the desired size from the same transfer function that was used for initializing the simulation;
    \item using the EFT coefficients fit from the simulation, paint on the full large-scale 21~cm field;
    \item to mitigate the effects of cosmic variance in making predictions for observables, repeat steps 3 and 4 in as many generated boxes as necessary.
\end{enumerate}
In this work, we focus on steps 1-4 and leave step 5 to future work.

The rest of this paper is dedicated to establishing the accuracy of this method for predicting the large-scale 21~cm field and is organized as follows. 
In Section~\ref{sec:formalism}, we review the relevant aspects of the formalism that we will use to relate the linear density field to the full 21~cm field. 
In Section~\ref{sec:methods}, we describe in detail how we can use the \textsc{thesan} simulations of reionization as a testing ground for our method. Specifically, we fit the EFT in a small sub-volume of the simulation box and use that to predict the rest of the simulation volume. 
In Section~\ref{sec:results}, we explore how variations in our method affect the accuracy of enlarging the simulations, finding that most of the predictive power can be encapsulated in a single coefficient and spectral shape. 
Concluding remarks follow in Section~\ref{sec:conclusion}.

\section{Formalism}
\label{sec:formalism}
The 21~cm brightness temperature $T_{21}$ can be expressed in terms of the matter overdensity $\delta$ as
\begin{align}
	 T_{21} &\approx \,28 (1+\delta) x_\mathrm{HI} \left(\frac{\Omega_b h^2}{0.0223}\right)\left(1 - \frac{T_\text{CMB} }{T_\text{spin}} \right) \nonumber\\ &\times \sqrt{\left(\frac{1+z}{10}\right) \left(\frac{0.24}{\Omega_m}\right)} \left(\frac{H(z)/(1+z)}{\mathrm{d}v_\parallel / \mathrm{d}r_\parallel}\right) \,\mathrm{mK} 
	\label{eqn:dTb}
\end{align}
where $x_\mathrm{HI}$ is the neutral hydrogen fraction, $\Omega_b$ and $\Omega_m$ are the baryon and matter densities in units of the critical density, $h$ is the Hubble parameter $H(z)$ at redshift $z=0$ in units of 100~km/s/Mpc, and $\mathrm{d}v_\parallel / \mathrm{d}r_\parallel$ is the line-of-sight proper motion gradient~\cite{Furlanetto:2019jso}.
In this work, we assume that we are sufficiently deep into the EoR that the spin temperature, which quantifies the relative occupancy of the spin-1 and spin-0 hyperfine states of hydrogen, is much larger than the temperature of the cosmic microwave background (CMB), $T_\text{spin}\gg T_\text{CMB}$, so that we can ignore local spin temperature fluctuations.

We then define $\delta_{21}$ as the fluctuations in the brightness temperature, $\delta_{21} = ( T_{21} - \avg{ T_{21}}) / \avg{ T_{21}}$. This will be the biased tracer field that we capture using EFT techniques. 
In particular, we will assume a bias expansion of the form 
\begin{equation}
    \delta_{21} = b_1 \delta + b_{\nabla^2} \frac{\nabla^2 \delta}{k_\mathrm{NL}^2} + b_2 \delta^2 + b_{\mathcal{G}_2} \mathcal{G}_2 + \ldots,
    \label{eq:bias}
\end{equation}
where the bias coefficients $b$ are dimensionless and $k_\mathrm{NL}$ is the wavenumber above which the field can no longer be treated perturbatively. Note that this expansion treats both $\delta$ and $k/k_\mathrm{NL}$ as perturbatively small quantities. 
In keeping with the EFT-based nomenclature, we refer to each of these terms as operators. 
Care must be taken when Fourier transforming the composite operators (i.e. terms involving products of fields) in this bias expansion, such as $\delta^2$ and the tidal operator $\mathcal{G}_2 = (\nabla_i \nabla_j \phi)(\nabla^i \nabla^j \phi) - \nabla^2 \phi$, which is expressed in terms of the gravitational potential satisfying the Poisson equation $\nabla^2\phi \sim \delta $. 
Notably, the position-space multiplication becomes a convolution over all Fourier-space modes, some of which are deep into the nonlinear regime where the density field cannot be modeled analytically. This means that the lack of theoretical control in the high-$k$ part of the integrand must be ``renormalized'' so that low-$k$ predictions are not affected by spurious high-$k$ contributions~\cite{Assassi:2014fva}. This renormalization can be done systematically, order by order. For instance, subtracting off UV-sensitive contributions to the Fourier transform of $\delta^2$ yields the renormalized $[\delta^2]$
\begin{equation}
    [\delta^2 ] = \delta^2 - \sigma^2(\Lambda)\Big(1 + \frac{68}{21}\delta + \frac{8126}{2205}\delta^2 + \frac{254}{2205} \mathcal{G}_2+ \ldots \Big) .
\end{equation}
Meanwhile, the tidal operator $\mathcal{G}_2$ does not need to be renormalized to leading order in $k/k_\text{NL}$~\cite{Assassi:2014fva}.

In order to then compute $\delta_{21}$ on large scales, one must determine the full non-linear density field to insert into the renormalized bias expansion. This is straightforward in the context of an $N$-body simulation where the density field is known, but if one wishes to only use information about the linear density field then it is possible to use techniques from standard perturbation theory (SPT) to compute the nonlinear density field. One can then also include contributions from the EFT of large-scale structure (LSS). In Fourier space, the density field can be expressed in terms of a perturbative ansatz, 
\begin{equation}
    \delta_{\boldsymbol{k}} = \sum_{n=1}^\infty \left(  a^n \delta^{(n)}_{\boldsymbol{k}} +  a^{n+2}\, \tilde{\delta}^{(n)}_{\boldsymbol{k}} \right) ,
\end{equation}
where $\delta^{(n)}$ and $\tilde{\delta}^{(n)}$ denote the $n$th-order density field and its EFT corrections, where the factors of $a^n$ arise from the linear growth factor in a matter-dominated universe, and where the EFT correction has an additional factor of $a^2$ so that the EFT terms have the same time-dependence as the loop diagrams from SPT. One can determine the $n$th-order densities by convolving $n$ copies of the linear density field with some convolution kernel, e.g. 
\begin{align}
    \delta^{(n)}_{\boldsymbol{k}} &= \int \dbar^3 q_1 \dots \int \dbar^3 q_n \, (2\pi)^3 \delta^D \left( \boldsymbol{k} - \sum_{i=1}^n \boldsymbol{q}_i \right) \nonumber \\ &\times F_n (\boldsymbol{q}_1, \dots, \boldsymbol{q}_n) \delta^{(1)}_{\boldsymbol{q}_1} \dots \delta^{(1)}_{\boldsymbol{q}_n} \\
    \tilde{\delta}^{(n)}_{\boldsymbol{k}} &= \int \dbar^3 q_1 \dots \int \dbar^3 q_n \, (2\pi)^3 \delta^D \left( \boldsymbol{k} - \sum_{i=1}^n q_i \right)  \nonumber\\&\tilde{F}_n (\boldsymbol{q}_1, \dots, \boldsymbol{q}_n) \delta^{(1)}_{\boldsymbol{q}_1} \dots \delta^{(1)}_{\boldsymbol{q}_n}.
\end{align}
The first few convolution kernels are 
\begin{align}
     F_1 &=  1 , \quad \tilde{F}_1 = -\frac{1}{9}c_s k^2 \label{eq:kernels} \\
    F_2(\boldsymbol{q}_1, \boldsymbol{q}_2) &= \frac{5}{7} + \frac{2}{7} \frac{(\boldsymbol{q}_1 \cdot \boldsymbol{q}_2)^2}{\boldsymbol{q}_1^2 \boldsymbol{q}_2^2} + \frac{\boldsymbol{q}_1 \cdot \boldsymbol{q}_2}{2} \left( \frac{1}{\boldsymbol{q}_1^2} + \frac{1}{\boldsymbol{q}_2^2} \right). \nonumber
\end{align}
Higher-order SPT kernels can be computed via well-known recursion relations~\cite{Bernardeau:2001qr,Goroff:1986ep,Jain:1993jh}, and the EFT kernels are compiled up to third order in Ref.~\cite{Bertolini:2016bmt}. The factor $c_s$ appearing in Eq.~\eqref{eq:kernels} is one of the EFT coefficients that has to be determined via a fit to simulation, and has the interpretation of an effective speed of sound for a self-gravitating fluid. However, we can see from the forms of Eqs.~\eqref{eq:kernels} and \eqref{eq:bias} that the $k$-dependence of the first EFT term and the scale-dependent linear bias term both scale as $k^2 \delta_k$, meaning that $c_s$ is completely degenerate with $b_{\nabla^2}$. We therefore omit $c_s$ from our fits because we work to leading order in perturbation theory, noting that unique (nondegenerate) EFT contributions to the density field would enter in a unique way at higher orders in perturbation theory. 

We finally note that the 21~cm brightness temperature is sensitive to the peculiar line-of-sight velocity and its gradient. This can be seen explicitly in Eq.~\eqref{eqn:dTb}, but also implicitly through the redshift dependence, since the observed redshifting of the 21~cm line will have a contribution from the peculiar velocity in addition to the expansion of the universe. Any measurement will therefore map the 21~cm field in ``redshift space'' coordinates $\boldsymbol{x}_r$ rather than in real space coordinates $\boldsymbol{x}$. Depending on the intended application of a simulation, it may be useful to paint on the modes of a biased tracer in redshift space rather than real space. 
It is straightforward to convert between the two via
\begin{equation}
	\boldsymbol{x}_r = \boldsymbol{x} + \frac{v_\parallel}
 {\mathcal{H}} \boldsymbol{\hat{n}} ,
\end{equation}
with $v_\parallel \equiv \boldsymbol{\hat{n}} \cdot \boldsymbol{v}_\mathrm{pec}$ where $\boldsymbol{\hat{n}}$ points along the line of sight in the simulation box and $\boldsymbol{v}_\mathrm{pec}$ is the peculiar bulk velocity of neutral hydrogen at the position $\boldsymbol{x}$. 
Note that at the redshifts of interest, this velocity should be the same as the matter velocity (since the origin of peculiar motion on large scales is the underlying gravitational field), which can be verified in the context of a particular simulation.
This has previously been shown to be correct at the percent level, see e.g. Refs.~\cite{2012MNRAS.422..926M,2011ApJ...730L...1S}. 
Using conservation of mass, one can relate real-space and redshift-space densities as $\delta_r (\boldsymbol{x}_r) = (1 + \delta (\boldsymbol{x})) \left| {\partial \boldsymbol{x}_r}/{\partial \boldsymbol{x}} \right|^{-1} -1 $, Fourier transform, and Taylor expand in the limit ${k_\parallel v_\parallel}/{\mathcal{H}} \ll 1$, where 
$k_\parallel \equiv \boldsymbol{\hat{n}} \cdot \boldsymbol{k}$. 
This leads to
\begin{align}
	(\delta_r)_{\boldsymbol{k}} =& \delta_{\boldsymbol{k}} 
	-i \frac{k_\parallel}{\mathcal{H}} (v_{\parallel})_{\boldsymbol{k}} 
	-i \frac{k_\parallel}{\mathcal{H}} \left({\delta v_{\parallel}}\right)_{\boldsymbol{k}}
	- \frac{1}{2} \left(\frac{k_\parallel}{\mathcal{H}} \right)^2 \left({v_\parallel^2}\right)_{\boldsymbol{k}} \n
	&- \frac{1}{2} \left(\frac{k_\parallel}{\mathcal{H}} \right)^2 \left({\delta v_\parallel^2}\right)_{\boldsymbol{k}}
	+ \frac{i}{6} \left(\frac{k_\parallel}{\mathcal{H}} \right)^3 \left({v_\parallel^3}\right)_{\boldsymbol{k}}
	+ \cdots.
	\label{eqn:RSD_expansion}
\end{align}
Combining the renormalized bias expansion for $\delta_{\boldsymbol{k}}$ with the mapping from real space to redshift space gives
\begin{align}
    (\delta_{21,r})_{\boldsymbol{k}} &= b_{1}^{(R)} \delta_{\boldsymbol{k}} - b_{\nabla^2} k^2 \delta_{\boldsymbol{k}} + b_{2}^{(R)} \left[ \delta^2 \right]_{\boldsymbol{k}} + b_{G2}^{(R)} (\mathcal{G}_2)_{\boldsymbol{k}} \n
    &-i \frac{k_\parallel}{\mathcal{H}} \Big[ (v_\parallel)_{\boldsymbol{k}} + b_1 (\delta v_\parallel)_{\boldsymbol{k}} - b_{\nabla^2} k^2 (\delta v_\parallel)_{\boldsymbol{k}} \Big]\nonumber\\& - \frac{1}{2} \left(\frac{k_\parallel}{\mathcal{H}} \right)^2 \left[ v_\parallel^2 \right]_{\boldsymbol{k}} + \cdots
    \label{eqn:d21_renorm}.
\end{align}
If desired, one can recover the real-space limit by setting $v_\parallel$ to zero. 

\section{Methods}
\label{sec:methods}

\subsection{Thesan simulations}
\label{sec:thesan}

%
\begin{figure}
    \includegraphics[width=\columnwidth]{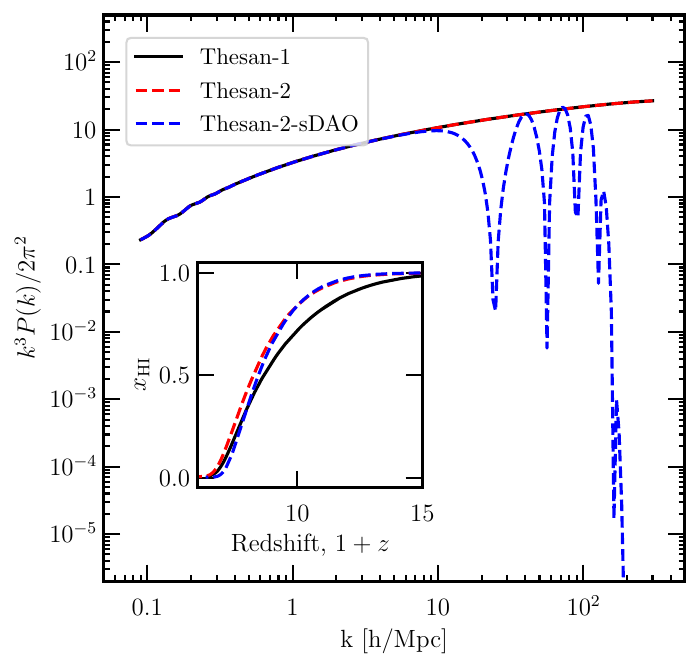}
    \vspace{-0.8cm}
    \caption{
    Power spectra for the initial density field in the \thesan simulations. \textsc{thesan-1} and \textsc{thesan-2} assume standard cold dark matter, while \textsc{thesan-2-sdao} includes dark acoustic oscillations.
    Inset shows the reionization history of the \thesan simulations.
    }
    \label{fig:transfer}
    \vspace{-0.3cm}
\end{figure}
The \thesan simulation suite is a set of cosmological radiation-magnetohydrodynamic simulations designed in particular for the study of reionization and high-redshift galaxy formation~\cite{thesan1}. 
Sub-resolution physics, including star formation, stellar feedback, and black hole accretion, is treated using the \textsc{IllustrisTNG} model of galaxy formation which has been shown to accurately match the observed properties of galaxies at low redshifts~\cite{2018MNRAS.475..624N,2018MNRAS.477.1206N,2018MNRAS.475..648P,2018MNRAS.480.5113M,2018MNRAS.475..676S}. 
The \thesan model for reionization uses the \textsc{arepo-rt} moving mesh hydrodynamic code to create self-consistent radiation transport, non-equilibrium heating and cooling, and realistic ionization sources that drive the ionization processes~\cite{2019MNRAS.485..117K}.
To capture dust dynamics, the dust is treated numerically as a property of the simulation's gas elements, with set prescriptions for its creation and destruction~\cite{2017MNRAS.468.1505M}. 
Combining these aspects, \thesan is able to accurately replicate observed properties of galaxies and the IGM at high redshifts~\cite{thesan2}. 

All \thesan simulations are made from boxes of comoving size 64.7 $h^{-1}$ Mpc (corresponding to 95.5 Mpc).
For the purposes of this work, we focus on the \textsc{thesan-1}, \textsc{thesan-2}, and \textsc{thesan-2-sdao} simulations, rendered on a $128\times128\times128$ grid, in order to characterize the distinguishability between reionization histories affected by different underlying physics. \textsc{thesan-1} is a high-resolution simulation containing $2100^3$ DM particles of mass $3.12 \times 10^6 \Msun$
and $2100^3$ gas particles of mass $5.82 \times 10^5 \Msun$. Meanwhile, both \textsc{thesan-2} and \textsc{thesan-2-sdao} simulations, which have a lower mass resolution compared to \textsc{thesan-1} by a factor of 8, contain $1050^3$ dark matter particles of mass $2.49 \times 10^7 \Msun$ and $1050^3$ gas particles of mass $4.66 \times 10^6 \Msun$. 
The primary difference between the simulations is that, while \textsc{thesan-1} and \textsc{thesan-2} follow standard $\Lambda$CDM cosmology, \textsc{thesan-2-sdao} assumes a transfer function that incorporates collisional damping from interactions between DM and dark radiation. 
As shown in Fig.~\ref{fig:transfer}, this results in a suppression of the matter power spectrum on small scales and the appearance of dark acoustic oscillations with the first peak at $k_\mathrm{peak}=$~40~$h^{-1}$Mpc. 
The inset of Fig.~\ref{fig:transfer} shows $x_\mathrm{HI}$ as a function of redshift for the different simulations.
Reionization for \textsc{thesan-1} begins earlier than the others, and the differences between \textsc{thesan-2} and \textsc{thesan-2-sdao} become more evident as reionization proceeds towards later redshifts, $1+z \lesssim 9$.

\subsection{Generating simulation subvolumes}
\label{sec:truncate}

Since the \textsc{thesan} data are available on a $128^3$ grid, it is straightforward to truncate the grid to an $N^3$ sub-volume for $N<128$ to represent a simulation with a smaller volume.
In principle, the operators/terms appearing in Eq.~\eqref{eqn:d21_renorm} should be recalculated for each value of $N$. 
However, the truncated boxes no longer obey the periodic boundary conditions imposed on the full box. Therefore, operators built from the density field in the truncated subvolume will contain unphysical artifacts due to the sharp edges at the boundaries.
Hence, we instead use the operators calculated from the linear density field present in the full simulation and pare them down to an $N^3$ grid. We emphasize that this step does not require information from the full simulation beyond the realization of the linear density field which is not computationally expensive to generate and follows straightforwardly from the initial conditions. 
We expect that the procedure would yield similar results as using operators calculated from a smaller simulation that \textit{does} have periodic boundary conditions within that volume.

\subsection{Principal component analysis}
\label{sec:PCA}

Previous works found that the operators/terms appearing in the bias expansion in Eq.~\eqref{eqn:d21_renorm} could be degenerate and have similar shapes in Fourier space, even when performing fits at the field level rather than fitting to a summary statistic like the power spectrum~\cite{21cmEFT}.
To determine the number of degrees of freedom that are actually well-constrained from fitting to simulations, we conduct a principal component analysis (PCA), which yields combinations of operators that have orthogonal impacts on the 21\,cm signal as well as an estimate of how well constrained each combination is.
Below, we briefly review the method of PCA.

The Fisher information is a measure of the information that an observable carries about a model (or more precisely, its parameters).
In this context, we can construct the Fisher information matrix as
\begin{equation}
    F_{ij} = 
    \frac{d (\delta_{21,r})_{\boldsymbol{k}}^*}{d b^{(R)}_i} 
    \, (\Sigma_{\boldsymbol{kq}})^{-1} 
    \, \frac{d (\delta_{21,r})_{\boldsymbol{q}}}{d b^{(R)}_j} ,
    \label{eq:fisher}
\end{equation}
where $\Sigma_{\boldsymbol{kq}}$ is the covariance matrix for the tracer field. 
Since Eq.~\eqref{eqn:d21_renorm} is linear in the bias coefficients, the derivatives reduce to the corresponding operators/terms, which we also smooth to ensure that we only include Fourier modes $k<k_\text{NL}$.
The principal components (PCs) are given by the normalized eigenvectors of the Fisher information matrix and are orthogonal so long as the eigenvalues are distinct, since the Fisher information matrix is symmetric by construction.
The first PC, which has the largest eigenvalue, corresponds to the combination of parameters that is the most well constrained, since it has the highest Fisher information.
PCs with smaller eigenvalues have a progressively smaller measurable impact on the observable.

For the method of fitting described in the next Subsection, each individual Fourier mode on scales $k<k_\text{NL}$ is given equal weight and assumed to be statistically independent of other modes.
In other words, the variance on each mode is assumed to be equal and the covariance between modes is zero. 
Hence, the covariance matrix corresponding to this set of assumptions is given in Fourier space by $\Sigma_{\boldsymbol{kq}} \propto \delta_{\boldsymbol{kq}}$, where $\delta_{\boldsymbol{kq}}$ here represents the Kronecker delta. 
The set of assumptions in choosing this covariance matrix is unlikely to be strictly correct; for instance, it is well known that non-linearities induce coupling between distinct Fourier modes, and even for a Gaussian field the variance is related to the power spectrum. While the former effect, which determines the off-diagonal components of the covariance matrix, is difficult to quantify, it is relatively straightforward to instead use a covariance matrix $\Sigma_{\boldsymbol{kq}} \propto P(k) \delta_{\boldsymbol{kq}}$. We have explicitly checked that using the power spectrum in the covariance matrix does not significantly alter our results, and below we show that the largest PCs determined using this approach accurately reproduce the full EFT fit. 
Therefore, our choice of covariance matrix does appear to appropriately capture the constraining power of the simulation in determining the principal components.

\subsection{Fitting coefficients at the field level}
\label{sec:fit}
As in Refs.~\cite{McQuinn:2018zwa,21cmEFT}, given a model of the 21\,cm signal, $\delta_\mathrm{EFT}$, which depends on some parameters such as the bias coefficients, we fit the model to simulations by minimizing the loss function
\begin{equation}
    \mathcal{A} = \sum_{k < k_\mathrm{NL}} P_\mathrm{err} (\boldsymbol{k}) = \sum_{k < k_\mathrm{NL}} \frac{|(\delta_\mathrm{sim})_{\boldsymbol{k}} - (\delta_\mathrm{EFT})_{\boldsymbol{k}}|^2}{V},
    \label{eqn:cost}
\end{equation}
where the sum is over all distinct wavevectors $\boldsymbol{k}$, $P_\mathrm{err} $ is the error power spectrum, and $V$ is the simulation volume. If we were to adopt a covariance matrix $\Sigma_{\boldsymbol{kq}} \propto P(k) \delta_{\boldsymbol{kq}}$ as discussed above, that would correspond to inverse-variance weighting the loss function with an additional factor of $P(k)$. In previous work, we used Eq.~\eqref{eqn:d21_renorm} as $\delta_\mathrm{EFT}$ to fit the bias parameters~\cite{21cmEFT}.
Equivalently, we can take $\delta_\mathrm{EFT}$ to be a linear combination of the principal components described in the previous section and thus fit for the coefficients of each principal component.
For the remainder of this work, we take the latter approach.

The loss function only includes modes with wavenumber less than $k_\mathrm{NL}$, the wavenumber above which we expect the bias expansion to break down as a valid descriptor of the 21\,cm signal.
We determine $k_\mathrm{NL}$ from the simulations by smoothing the 21\,cm field until the relative fluctuations take values less than 0.8~\cite{21cmEFT}.
For example, for \textsc{thesan-1} at a redshift of $z=8.5$, this corresponds to $k_\text{NL} = 0.4$\,$h$ Mpc$^{-1}$.
We have explicitly checked that our results do not depend sensitively on the choice of 0.8 as the maximum fluctuation size, as they do not change substantially when $k_\mathrm{NL}$ is varied by $\Delta k_\mathrm{NL} \sim 0.1 \,h$ Mpc$^{-1}$.
The value of $k_\mathrm{NL}$ also sets the smallest possible simulation volume to which our ``supersizing" procedure can be applied, since we expect the effective field theory description to completely fail for boxes smaller than about $2 \pi / k_\mathrm{NL} = 16$ $h^{-1}$ Mpc on each side at this redshift due to a lack of perturbative modes.



\begin{figure}
    \includegraphics[width=\columnwidth]{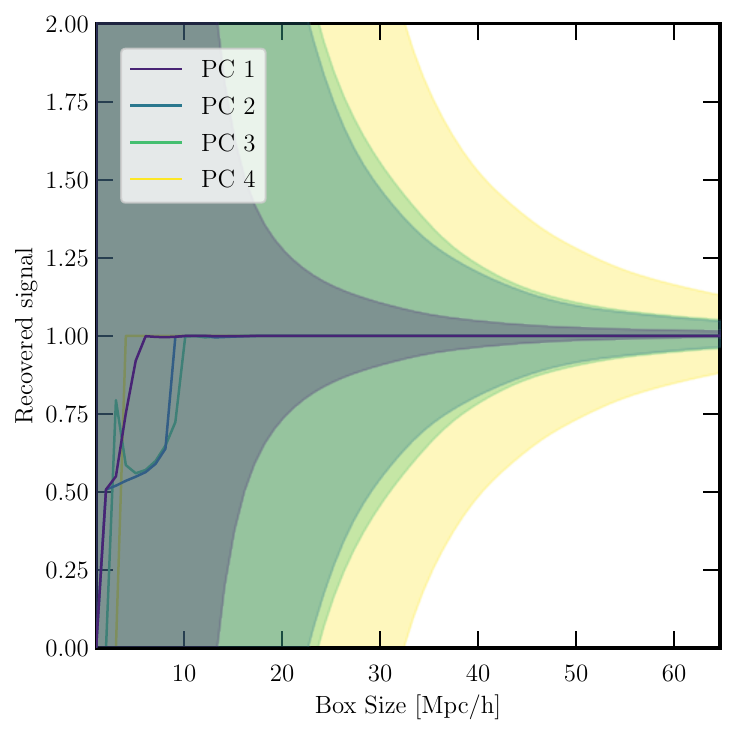}

    \caption{
        The recovered coefficient of the PCs, when fitting to a field constructed from a single PC (normalized to unity).
        For each injection, we obtain the correct PC with a coefficient of unity, to well within fitting uncertainties---the coefficients of PCs that were not injected are found to be consistent with zero and are not shown here.
    }
    \label{fig:sig_inj}
\end{figure}

\subsection{Validation}
As a validation of the methods described in this Section, we perform a signal-injection-like test to ensure that we correctly recover the coefficients of the PCs, or equivalently, the bias coefficients.
For each subvolume size, we construct a field that consists of a single principal component with its coefficient set to unity, and use the minimization procedure described in Sec.~\ref{sec:fit} to see if we recover the this principal component.

Fig.~\ref{fig:sig_inj} shows the level of recovery for each principal component, along with the 68\% confidence intervals.
For each injected signal, the correct PC is recovered with a coefficient of unity down to a box size of about 10 Mpc/h.
The coefficients of the non-injected PCs are not shown, as we find they are always equal to zero well within the level of uncertainty.
In addition, as is expected from the results of the PCA, the uncertainty on the recovered signal is smallest for the first PC, and increases with each PC of subsequently smaller eigenvalue.

\section{Results}
\label{sec:results}

In order to establish whether perturbative methods can be used to predict the super-sample modes that would be obtained with a larger hydrodynamical simulation, we must assess whether the 21\,cm differential brightness temperature of the full \thesan simulation can be correctly inferred from a subvolume of the simulation.
Starting with the full \thesan simulations on a $128^3$ grid, we truncate the simulation boxes to an $N^3$ grid for all values of $0 < N < 128$, run a PCA, then fit the coefficients of the PCs to the truncated simulation.
We begin in Section~\ref{sec:firstPC} by discussing the dominance of the first principal component and the stability of the first PC across different values of $N$, before showing the fits to simulation subvolumes in Section~\ref{sec:subvolumes}.

\subsection{The first principal component}
\label{sec:firstPC}
%
\begin{figure}
    \includegraphics[width=\columnwidth]{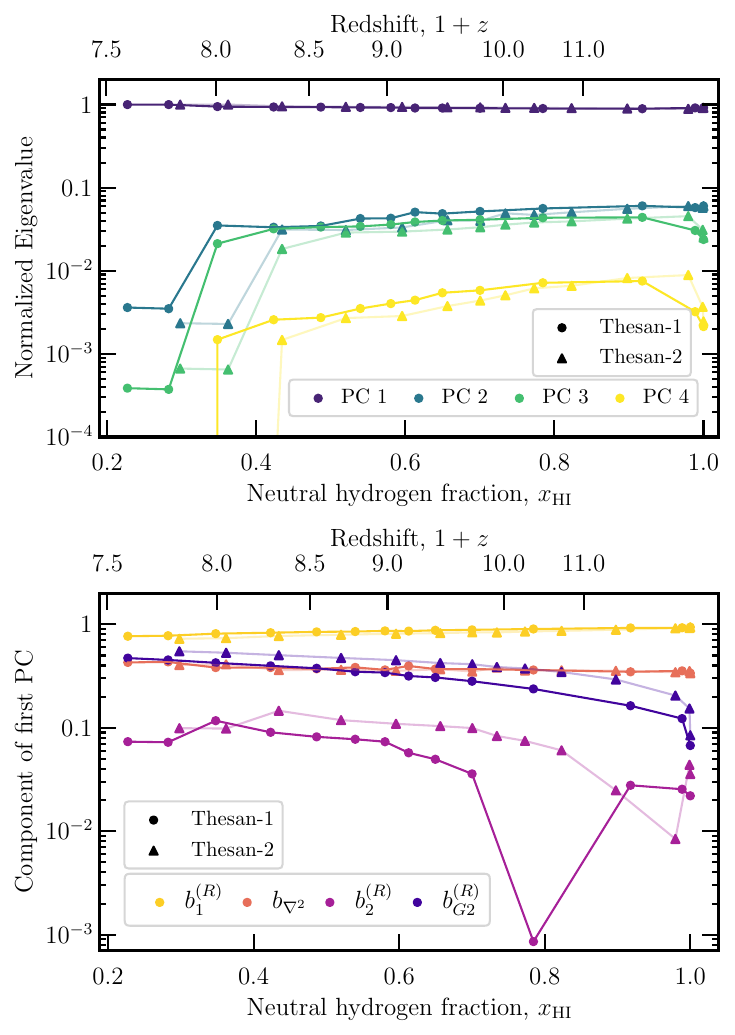}
    \vspace{-0.6cm}
    \caption{
        Eigenvalues of the Fisher matrix (top) and EFT-operator contributions to the first principal component (bottom) for \textsc{thesan-1} and \textsc{thesan-2} as a function of $x_\mathrm{HI}$. 
        The corresponding curves for \textsc{thesan-2-sdao} differ only at the few percent level from \textsc{thesan-2} and are visually indistinguishable. 
        In the lower panel, all components are postive except for $b_{\nabla^2}$, which is negative across all redshifts and for both simulations.
    }
    \label{fig:xHI}
    \vspace{-0.2cm}
\end{figure}
\begin{figure}
    \includegraphics[width=\columnwidth]{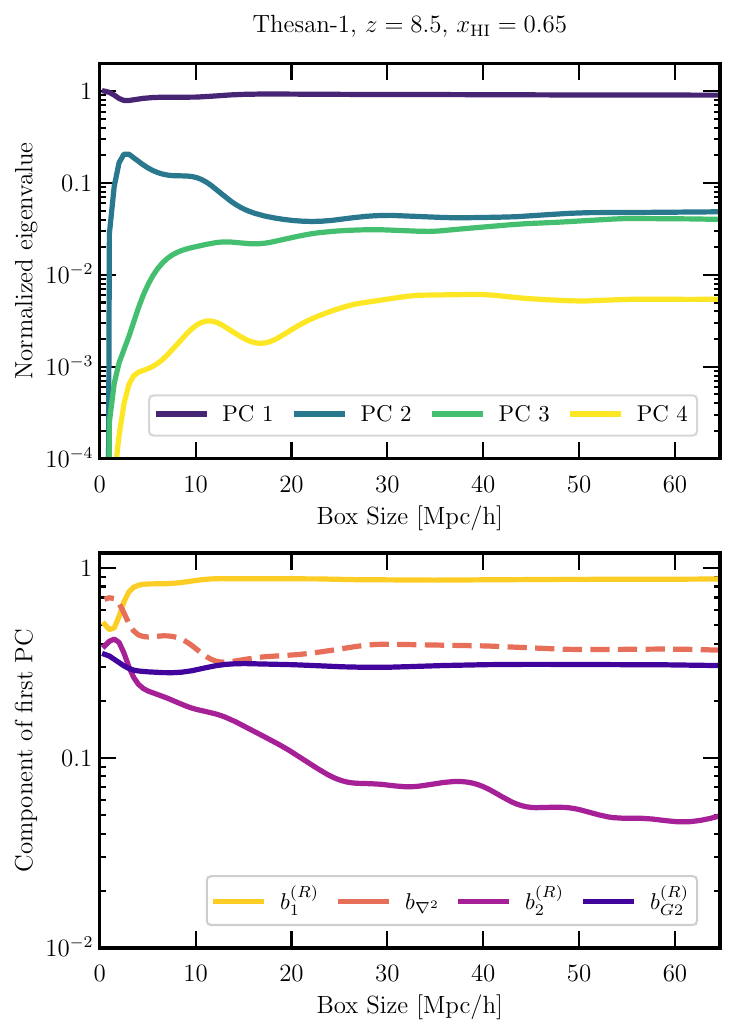}
    \caption{
        Eigenvalues of the Fisher matrix (top) and EFT-operator contributions to the first principal component (bottom) determined from subvolumes of \textsc{thesan-1}. For boxes larger than about 20 Mpc/h, the principal components converge to their full-box behavior. 
    }
    \label{fig:PCs}
\end{figure}

In Fig.~\ref{fig:xHI}, we show properties of the principal components for \textsc{thesan-1} and \textsc{thesan-2} as a function of $x_\mathrm{HI}$ and redshift.
The top panel shows how the eigenvalues of the Fisher matrix corresponding to different principal components (determined using the full simulation volume) vary across different values of $x_\mathrm{HI}$.
The eigenvalues are normalized such that the sum of the eigenvalues adds up to one.
For all simulations, the first principal component always comprises at least $88 \%$ of the Fisher information, demonstrating that the 21\,cm signal is well characterized by a single degree of freedom in the bias expansion.
For most of reionization, the second and third principal components together comprise about $10\%$ of the variation in the Fisher matrix and hence are non-negligible contributions; the fourth principal component, on the other hand, has an eigenvalue that is always less than one percent. 

These results suggest that although the bias expansion contains four degrees of freedom corresponding to the different bias parameters, only one degree of freedom can be constrained with great precision even when fitting simulations at the field level, regardless of which of the \thesan simulations we consider. 
Notably, the same renormalized bias expansion in Eq.~\eqref{eqn:d21_renorm} applies to any biased tracer of the matter field, with differences between tracers arising primarily due to different bias coefficients. 
Since the determination of the principal components is independent of the bias coefficients, this suggests that a similar result (i.e. the dominance of one principal component) will hold for other biased tracers beyond just the 21~cm field. 
The operators/terms appearing in Eq.~\eqref{eqn:d21_renorm} do have some mild dependence on the fact that we are considering the 21~cm field through the dependence on $k_\text{NL}$, which is determined by smoothing $\delta_{21}$ as described in Section~\ref{sec:fit}. 
This mild dependence on $k_\text{NL}$ can be seen in Fig.~\ref{fig:xHI}, where the eigenvalues of the less constrained principal components decrease as reionization proceeds. 
As $x_\mathrm{HI}$ drops below about 0.4 and $k_\mathrm{NL}$ decreases, the 21\,cm signal eventually drops out of the regime of the EFT's validity and the eigenvalues show much greater variation~\cite{21cmEFT}. 

The bottom panel of Fig.~\ref{fig:xHI} shows the composition of the first principal component.
For most of the duration of the simulation, the first principal components of \textsc{thesan-1} and \textsc{thesan-2} are primarily comprised of $b_1^{(R)}$, the linear bias coefficient, with significant contributions from $b_{\nabla^2}$, which is closely related to the size of ionized bubbles~\cite{McQuinn:2018zwa,21cmEFT}, and $b_{G2}^{(R)}$, which represents contributions from anisotropic stress or tidal forces.

\begin{figure*}
    \includegraphics[width=\textwidth]{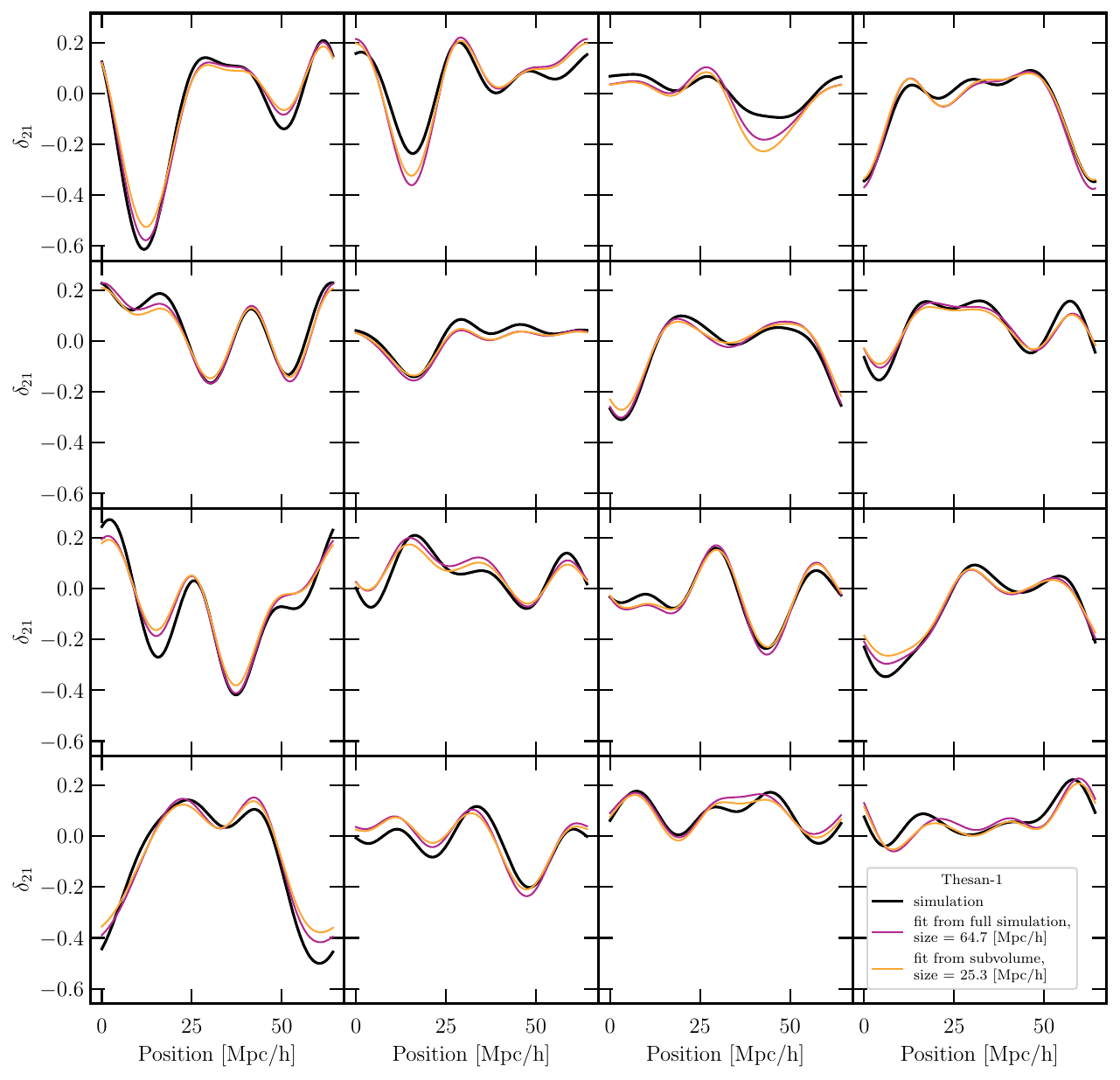}
    \caption{
    Real-space fluctuations in the 21\,cm differential brightness temperature along evenly spaced slices through \textsc{thesan-1}. 
    Also shown are the EFT predictions in those same slices as determined by fitting all four principal components at the field level in different simulation volumes. 
    The EFT reproduces the simulation with relative differences at the level of $\mathcal{O} (10\%)$, and the prediction is relatively insensitive to whether we use the full simulation to fit the bias coefficients or a $\sim5$\% subvolume.
    }
    \label{fig:1D_slices}
\end{figure*}

Focusing on the \textsc{thesan-1} simulation at a particular snapshot in time, with a mean free electron fraction of $x_\mathrm{HI} = n_\mathrm{HI} / n_H = 0.65$ corresponding to a redshift of $z = 8.5$, the top panel of Fig.~\ref{fig:PCs} shows the normalized eigenvalues of the Fisher matrix determined from different simulation subvolumes.
The eigenvalues corresponding to all the PCs are quite stable across different values of $N$, or, equivalently, the side length of the simulation box, and only begin to show some variability for side lengths of less than about 25 Mpc/h, corresponding to $N \lesssim 50$. 
This variability is to be expected, as the behavior corresponds to when the box size approaches $2\pi/k_\mathrm{NL}$, and hence there are few modes within the simulation available to fit.

Moreover, the PCs are fairly stable across different subvolume sizes.
The bottom panel of Fig.~\ref{fig:PCs} shows the components of the first principal component for different subvolume side lengths.
Again, we see that for \textsc{thesan-1}, the first component is mostly comprised of $b_1^{(R)}$ across almost all subvolume sizes.
The $b_1^{(R)}$, $b_{\nabla^2}$, and $b_{G2}^{(R)}$ components are nearly constant across the entire range, with $b_2^{(R)}$ showing slight variability.
The components begin to fluctuate significantly below a simulation box size of about 25 Mpc/h, similar to what can be seen in the top panel of Fig.~\ref{fig:PCs}. 
The stability of the PCs and their eigenvalues even down to fairly small subvolumes is a promising indication that the observables of a large simulation can be captured from a smaller simulation run with the same physical parameters.
In the next section, we show this explicitly in the field-level 21\,cm signal.

\subsection{Fitting subvolumes}
\label{sec:subvolumes}

Fig.~\ref{fig:1D_slices} shows several evenly spaced slices of the 21~cm field predicted by the EFT, which includes all four of the principal components in the fit, compared to the true 21\,cm signal from \textsc{thesan-1}. 
The predicted 21~cm field is relatively robust to changes in the simulation volume over which the best-fit parameters are determined. In other words, even when we truncate the box to a side length of $25.3$ Mpc/h, any change to the best-fit EFT coefficients (as compared to the full-volume best-fit coefficients) yields differences at the field level that are not visually significant.
This demonstrates that the 21\,cm differential brightness temperature in \textsc{thesan-1} can be predicted at the field level on large scales even from a simulation that represents only $\sim$5\% of the volume of \textsc{thesan-1}, although the exact size threshold likely depends on other parameters such as $k_\mathrm{NL}$ and the grid spacing. 

\begin{figure}
    \includegraphics[width=\columnwidth]{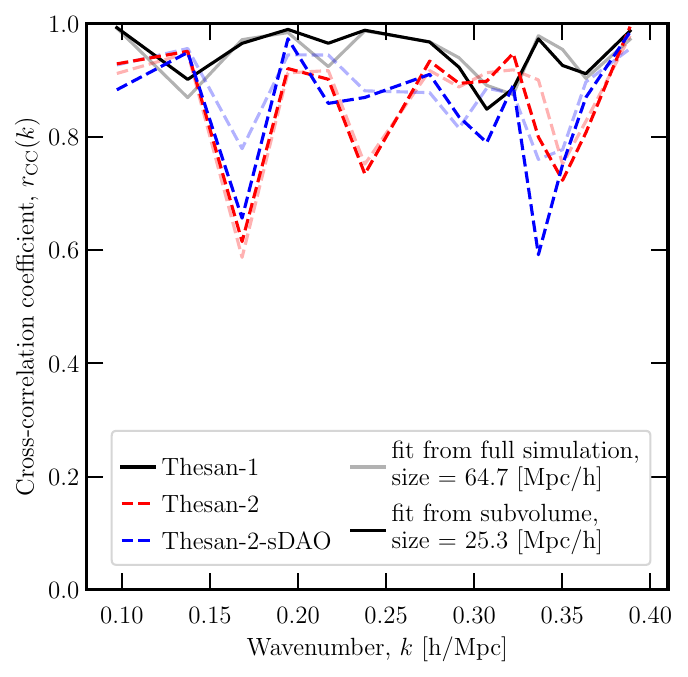}
    \caption{
    Cross-correlation coefficient, $r_\mathrm{CC} (k)$ for wavenumbers up to $k_\mathrm{NL}$.
    }
    \label{fig:CCC}
\end{figure}
To better quantify the agreement of the EFT bias expansion with the simulations, we use the cross-correlation coefficient defined as
\begin{equation}
    r_\mathrm{CC} (\mathbf{k}) = \frac{P_{XY} (\mathbf{k})}{\sqrt{P_X (\mathbf{k}) P_Y (\mathbf{k})}} .
\end{equation}
The cross-correlation is shown in Fig.~\ref{fig:CCC} for \textsc{thesan-1}, \textsc{thesan-2} and \textsc{thesan-2-sdao}.
We find that the fits determined from both the full simulation and the $(25.3 \,\mathrm{Mpc/h})^3$ subvolume reproduce the simulation with $1 - r_\mathrm{CC} (k < k_\mathrm{NL}) \sim 10\%$, consistent with the level of agreement seen in previous work~\cite{21cmEFT}. 

\begin{figure}
    \includegraphics[width=\columnwidth]{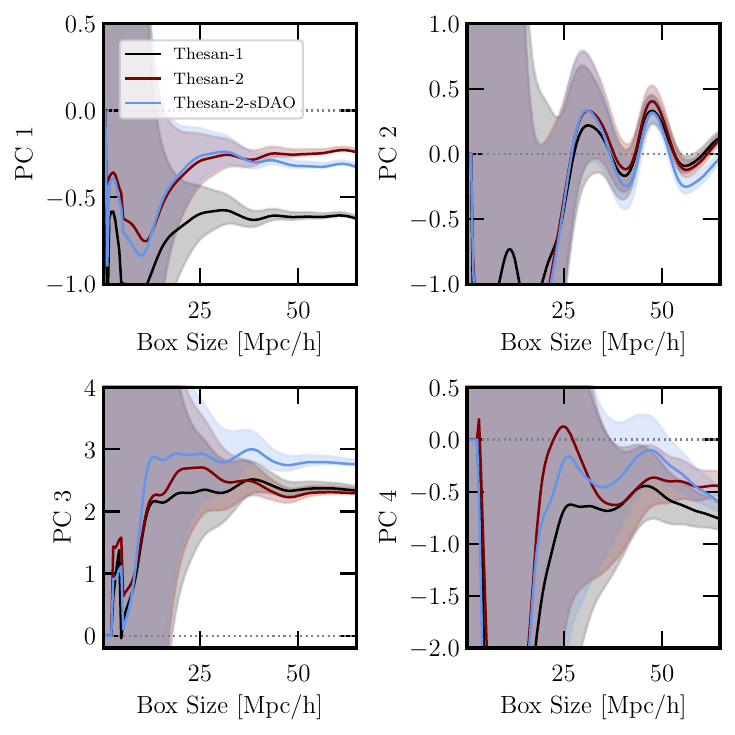}
    \caption{
        Coefficients of the principal components fit using different subvolumes of the simulation. The bands show the 68\% confidence intervals. For $N\lesssim50$, corresponding to subvolumes smaller than $(25.3 \,\mathrm{Mpc/h})^3$, there are too few modes with $k < k_\mathrm{NL}$ to provide a good fit. 
    }
    \label{fig:coeffs_PC}
\end{figure}

\begin{figure}
    \includegraphics[width=\columnwidth]{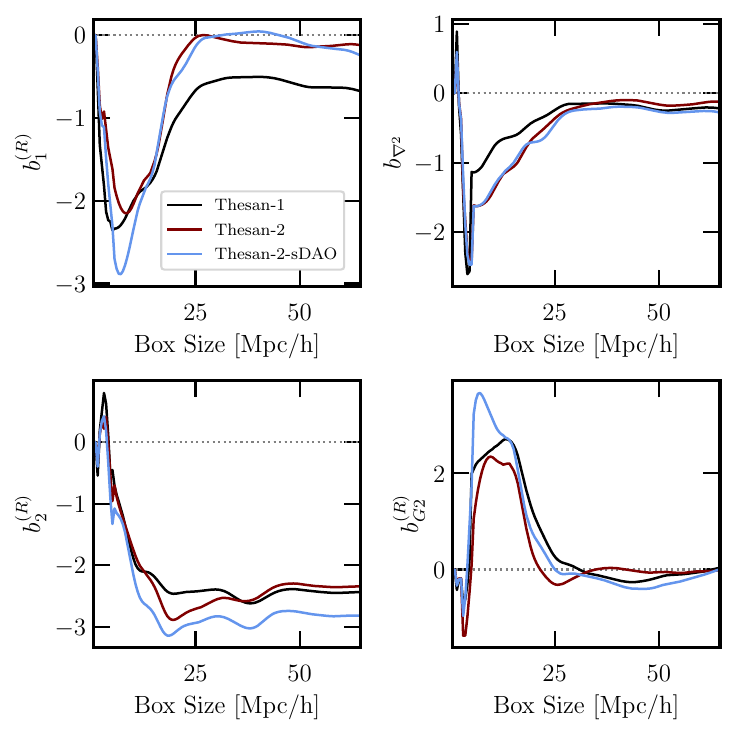}
    \caption{Bias coefficients corresponding to the best fits in Fig.~\ref{fig:coeffs_PC}.
    All four coefficients are stable down to box sizes of about $25$ Mpc/h.}
    \label{fig:coeffs_bias}
\end{figure}

Interestingly, the agreement between the EFT and simulation is best for the highest-resolution simulation we consider, \textsc{thesan-1}. 
This may be due to the relative importance of nonlinear terms in the \textsc{thesan-2} simulations, as demonstrated in Figs.~\ref{fig:coeffs_PC} and \ref{fig:coeffs_bias} where
we show the best-fit principal component coefficients and bias coefficients for \textsc{thesan-1}, \textsc{thesan-2}, and \textsc{thesan-2-sdao}.
The uncertainties are such that only the first and third PCs have coefficients that are distinguishable between the different simulations at a significant level.
The first PC appears to encapsulate physics primarily related to the mass resolution of each simulation, whereas the third PC seems to capture information about the underlying dark matter physics. While this behaviour is easiest to see for the largest simulation volumes we fit to, this information is preserved even for small simulations. For simulation box sizes larger than $(25.3 \,\mathrm{Mpc}/h)^3$, the coefficients of the first and third PCs vary by less than 10\% around their respective mean values for both simulations; this, combined with the size of the uncertainties, indicates that these coefficients are reliably distinguishable from zero across a large range of box sizes. 
In contrast, the coefficients of the second and fourth PCs are not significantly different between the three simulations. They also vary substantially, even changing sign across different subvolume sizes, making their extraction and interpretation less reliable.

\begin{figure*}
    \centering
    \includegraphics[width=\textwidth]{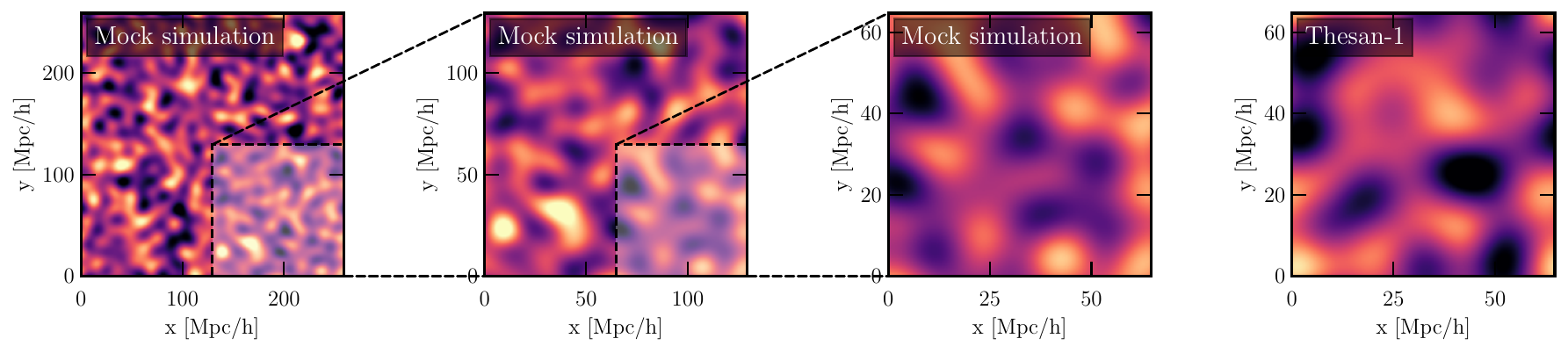}
    \caption{  
        A demonstration of the ``supersizing" procedure. 
        The left panel shows the 21\,cm EFT prediction, smoothed over $k > k_\mathrm{NL}$, generated using bias coefficients fit from a subvolume of\thesan-1 and applied to a density field with a volume that is $4^3$ times larger than the \thesan volume. The second panel shows a zoom-in of a section that is $2^3$ times the volume of \thesan, and the third panel shows a zoom-in of a section that is the same volume.
        In the right panel, we show the \thesan-1 simulation smoothed on scales smaller than $k_\mathrm{NL}$ for comparison.
    }
    \label{fig:big_slice}
\end{figure*}
Summing the principal components and their coefficients gives the best-fit bias parameters, which are shown in Fig.~\ref{fig:coeffs_bias}.
For all simulations, the first PC is mainly comprised of $b_1^{(R)}$, so the most constrainable contribution to the 21\,cm signal is the linear renormalized bias. Interestingly, the inferred linear bias is close to zero for the \textsc{thesan-2} and \textsc{thesan-2-sdao} simulations, which may explain why the EFT performs slightly worse at reproducing these simulations compared with \textsc{thesan-1} as quantified by $r_\mathrm{CC}$. 
The third PC, which is the only other PC that is distinguishable between the simulations, is dominated by quadratic bias, $b_2^{(R)}$.
This appears to be consistent with previous field-level fits to the \thesan-2 suite, which found that $b_2^{(R)}$ was the coefficient with the largest magnitude (particularly after nondimensionalizing $b_{\nabla^2}$ with factors of $k_\text{NL}$)~\cite{21cmEFT}. Moreover, as is discussed at length in Ref.~\cite{McQuinn:2018zwa}, $b_2^{(R)}$ is physically related to the clustering of the sources of ionizing radiation. Therefore, it is perhaps to be expected that $b_2^{(R)}$ differs between simulations with different small-scale matter power spectra due to the underlying dark matter physics, as shown in Fig.~\ref{fig:transfer}.
Of all the EFT coefficients, the inclusion of $b_2^{(R)}$ was shown to have the largest impact on reducing the error power spectrum (see Fig.~5 of Ref.~\cite{McQuinn:2018zwa}) due to the patchiness of reionization. 

\subsection{Supersized simulation}

In Fig.~\ref{fig:big_slice}, we provide a demonstration of how our ``supersizing" prescription works in practice. 
We generate a large-scale linear density field that is 64 times larger in volume than the \thesan simulations and apply the effective bias expansion with coefficients fit from a volume representing just $\sim5$\% of the full-sized \textsc{thesan-1} simulation.
As was done for the initial conditions of \thesan, we generate the initial density field such that the amplitude of the Fourier modes is fixed to the ensemble power spectrum in order to mitigate cosmic variance.
To further reduce the impact of cosmic variance on observables of interest, one can apply this procedure to several density fields generated this way, or to another realization that is exactly out of phase~\cite{Giri:2022ijp}, but we leave exploration of this step to future work.
The leftmost panel shows a cross-section of the 21\,cm signal from the large mock simulation smoothed over $k>k_\mathrm{NL}$, while second and third panels show zoomed-in sections of the mock simulation.
The third panel is visually quite similar to \textsc{thesan-1}, shown in the last panel, which verifies that our procedure can generate large ``simulations" that reproduce the properties of the smaller simulation.

\section{Conclusion}
\label{sec:conclusion}

In this work, we have demonstrated the utility of EFT-inspired techniques for ``supersizing" simulations of the 21\,cm differential brightness temperature using the \thesan suite of simulations as a testbed.
As a proxy for simulations run with smaller volumes, we truncate the simulation boxes to a smaller size, perform a principal component analysis to identify how many degrees of freedom are necessary to accurately describe the 21\,cm signal, and fit the principal components to the simulation field. 
The principal components that are the most well constrained are primarily related to $b_1^{(R)}$, the linear bias, and $b_2^{(R)}$, which is related to source clustering.
When fitting the PCs to the simulations, we find that if the box is larger than $\sim 2 \pi / k_\mathrm{NL}$, one can accurately reproduce the large-scale 21\,cm fluctuations.

In addition, we find that our method is able to distinguish between simulations run with different underlying assumptions about dark matter physics and sub-grid modeling (due to e.g. the resolution effects that differ between \textsc{thesan-1} and \textsc{thesan-2}).
When performing fits on a simulation with a CDM initial matter power spectrum and a simulation that includes dark acoustic oscillations, the value of $b_2^{(R)}$ is statistically distinguishable between the two. 
We anticipate that this method will have the power to differentiate between other scenarios that alter the linear and quadratic bias on large scales. 
Although we focus on comparisons to the \thesan simulation suite in this work, one could also explore whether these conclusions hold when using other simulations such as those from Refs.~\cite{ciardi2003simulating,iliev2006simulating,trac2007radiative,gnedin2014cosmic,kaurov2016cosmic,pawlik2017aurora}, 
or study how a wider range of reionization morphologies would impact the value of the quadratic bias using the simulations from Refs.~\cite{McQuinn:2006et,Cain:2022mhv}. 

Compared to the \thesan simulations, which required nearly 60,000 cores and 30 million CPU hours to complete~\cite{top500-2020,thesan1}, the PCA and effective bias expansion 
can be calculated in minutes on a laptop, highlighting the immense computing resources that can be compressed by the EFT-based method. To produce the mock simulations shown in Fig.~\ref{fig:big_slice} necessitates the use of modest computational resources on a cluster, which is primarily driven by memory usage rather than by the need to perform many computations in parallel. 
Hence, our method will facilitate comparison between simulations and large-scale observations without needing to assume that structure formation proceeds according to linear theory or using a semi-analytic model like 21cmFAST, as was done in previous work~\cite{2022MNRAS.516.5588G}.
Our EFT-based method is also not necessarily limited to 21\,cm cosmology and could be applied to simulations of other biased tracers like the Lyman-$\alpha$ forest.
We leave this as a topic for future exploration, and anticipate that our procedure will be useful for other applications where it is necessary both to capture the physics of small scales while also aggregating sufficiently many modes on large scales at the field level.

\section*{Acknowledgments}

It is a pleasure to thank Adrian Liu, Matt McQuinn, Chirag Modi, Julian Mu\~noz, and Aaron Smith for useful conversations and correspondence pertaining to this work.
W.Q. was supported by the National Science Foundation Graduate Research Fellowship under Grant No. 2141064 and a grant from the Simons Foundation (Grant Number SFI-MPS-SFJ-00006250, W.Q.).
K.S. acknowledges support from a Natural Sciences and
Engineering Research Council of Canada Subatomic Physics Discovery Grant and from the Canada Research Chairs program. 
O.R. was supported by MIT's Undergraduate Research Opportunities Program (UROP) through the Paul E. Gray (1954) Fund and the Reed Fund.
S.O. was supported by the National Science Foundation under Grant No. AST-2307787.
This analysis made use of \texttt{Numpy} \cite{Harris:2020xlr}, \texttt{Scipy} \cite{2020NatMe..17..261V}, \texttt{Jupyter}~\cite{Kluyver2016JupyterN}, \texttt{tqdm}~\cite{daCosta-Luis2019}, and \texttt{Matplotlib} \cite{Hunter:2007ouj}.


\bibliography{main}
\end{document}